\documentstyle[preprint, epsf,aps]{revtex}
\newcommand{\laa}{{\cal {L}}_a(t)}
\newcommand{\cald}{\cal {D}}
\begin{document}
\draft
\title{A Geometric Fractal Growth Model for Scale Free Networks\\}
\author{S. Jung$^1$, S. Kim$^1$ and B. Kahng$^2$}
\address{$^1$ Nonlinear and Complex Systems Laboratory, Department 
of Physics, Pohang University of Science and Technology, Pohang, 
Kyongbuk 790-784, Korea\\
$^2$ School of Physics and Center for Theoretical Physics, Seoul 
National University, Seoul 151-747, Korea\\}
\date{\today}
\maketitle
\begin{abstract}
We introduce a deterministic model for scale-free networks, whose 
degree distribution follows a power-law with an exponent $\gamma$. 
At each time step, each vertex generates its offsprings, whose 
number is proportional to the degree of that vertex with proportionality 
constant $m-1$ ($m >1$). We consider the two cases: first, each 
offspring is connected to its parent vertex only, forming a tree 
structure, and secondly, it is connected to both its parent and 
grandparent vertices, forming a loop structure. We find that both models 
exhibit power-law behaviors in their degree distributions with the 
exponent $\gamma=1+\ln (2m-1)/\ln m$. Thus, by tuning $m$, the 
degree exponent can be adjusted in the range, $2 <\gamma < 3$. 
We also solve analytically a mean shortest-path distance $d$ between 
two vertices for the tree structure, showing the small-world behavior, 
that is, $d\sim \ln N/\ln {\bar k}$, where $N$ is system size, 
and $\bar k$ is the mean degree. Finally, we consider the case 
that the number of offsprings is the same for all vertices, 
and find that the degree distribution exhibits an 
exponential-decay behavior.    
\end{abstract}

\section{Introduction}
Recently, complex systems have received considerable attention as 
an interdisciplinary subject\cite{ab,mendes}. Complex systems consist 
of many constituents such as individuals, substrates, and companies in 
social, biological, and economic systems, respectively, 
showing cooperative phenomena between constituents through 
diverse interactions and adaptations to the pattern they 
create\cite{nature,science}. 
Recently, there have been a lot of efforts to understand such 
complex systems in terms of networks, composed of vertices and 
edges, where vertices (edges) represent constituents (their interactions).
This approach was initiated by Erd\"os and R\'enyi (ER)\cite{er}. 
In the ER model, the number of vertices is fixed, while edges connecting 
one vertex to another occur randomly with a certain probability. 
However, the ER model is too random to describe real complex systems. 
Recently, Barab\'asi and Albert (BA)\cite{ba,physica} introduced 
an evolving network where the number of vertices $N$ increases 
linearly with time rather than fixed, and a newly born vertex 
is connected to already existing vertices, following the 
so-called preferential attachment (PA) rule; 
When the number of edges $k$ incident upon a vertex is called the 
degree of the vertex, the PA rule means that the probability $\Pi_i$ 
for the new vertex to connect to an already existing vertex $i$ is 
proportional to the degree $k_i$ of the selected vertex, that is,  
\begin{equation}
\Pi_i = \frac{k_i}{\sum_j k_j}.
\label{pa}
\end{equation}
The main difference between the ER and BA models appears in  
the degree distribution. For the ER network, the degree distribution 
follows the Poisson distribution, while for the BA network, it follows 
a power-law, $P(k)\sim k^{-\gamma}$ with $\gamma=3$. 
The network whose degree distribution follows a power-law is called 
the scale-free (SF) network\cite{physica}. 
SF networks are abundant in real-world such as the world-wide 
web\cite{www1,www2,www3,www4}, the Internet
\cite{internet1,internet2,internet3,internet4}, 
the citation network\cite{redner}, the author collaboration 
network of scientific papers\cite{coworker}, and the metabolic 
networks in biological organisms\cite{metabolic}.\\ 

While a lot of models have been introduced to describe SF networks 
in real world, most of them are stochastic models. However, a couple 
of models recently introduced by Barab\'asi, Ravasz and Vicsek 
(BRV)\cite{br}, and Dorogovtsev and Mendes (DM)\cite{mendes} 
are deterministic. 
In general, the deterministic model is useful in investigating analytically 
not only topological features of SF networks in detail, but also 
dynamical problems on the networks. Both the BRV and the DM models are 
meaningful since they are not only the first attempts for deterministic 
SF networks, but also the ones constructed in a hierarchical way, 
so that analytic treatments can be made easily using recursive 
relations derived from two structures in successive generations. 
In the BRV model, however, the mean shortest-path distance between 
two vertices, called the diameter, is independent of system size. 
Thus, the model may be relevant to some specific systems 
such as the metabolic network\cite{metabolic}, where the diameter is 
independent of system size. In this paper, we introduce another type 
of the deterministic model for the SF network, which is also constructed 
in a hierarchical way. Our model is based on almost the same idea as that 
of the DM model. While the DM model starts from a triangle, our model 
does from a tree structure. This difference makes one easily modify 
the model into more general cases such as loopless or loop structures, 
and the ones with a various number of branches. Moreover, the simplicity of 
our model enables us to obtain the analytic solution for the degree 
distribution and the diameter. In particular, our model includes a 
control parameter, so that by 
tuning the parameter, we can obtain SF networks with a variety of 
degree exponents in the range, $2 < \gamma < 3$. Therefore our model 
should be useful to represent various SF networks in real world.\\

This paper is organized as follows. In section II, we will introduce 
deterministic models specifically for tree and loop structures, 
respectively. In section III, analytic treatment will be performed 
for the deterministic models introduced in section II. 
The final section will be devoted to conclusions and discussions. 

\section{deterministic model}
\label{sec1}
It is known that the number of vertices in most of SF networks in 
real world increases exponentially in time. Thus, our deterministic 
model is constructed in an evolving way, where each already existing 
vertex produces its offsprings, and the connections are made between 
old and new vertices. Thus, vertices are generated in a hierarchical 
order, so that the number of vertices increases geometrically in time.  
On the other hand, it is known \cite{rate1,rate2} that the PA probability 
$\Pi'_i$, Eq.(\ref{pa}) is generalized for real networks as 
\begin{equation}
\Pi'_i=\frac{k_i+\mu}{\sum_j k_j+\mu}, 
\end{equation}
where $\mu$ accounts for some randomness in connecting edges. 
To take into account of this modified PA behavior, we introduce 
two rules, called the addition and the multiplication rule, in the 
deterministic model, depending on how new vertices are generated 
from each old vertex. The details on both rules will be described below.   

\subsection{Tree structure}
The network forms a tree structure when new vertices generated from an 
old vertex are connected to their parent only. 

\subsubsection{The addition rule}

In the case of the addition rule, at each time step, a constant number of new 
vertices, say $\ell$ new vertices, are generated from each already 
existing vertex, and they are connected to their parent only. 
Then the degree $k_{i,a}$ at vertex $i$, where the subscript $a$ 
means the addition rule, evolves as 
\begin{equation}
k_{i,a}(t+1)=k_{i,a}(t)+\ell,  
\label{a_rate}
\end{equation}
so that 
\begin{equation}
k_{i,a}(t)=1+\ell(t-t_i),
\label{kia}
\end{equation} 
for $t \ge t_i$, where $t_i$ means the time when the vertex $i$ was born. 
Then the total number of vertices newly born at time $t$ becomes 
$\laa=\ell(1+\ell)^t$ for $t\ge 1$, and the total number of vertices 
$N_a(t)$ present at time $t$ is 
\begin{equation}
N_a(t)=\sum_{j=0}^t {\cal {L}}_a(j)=(1+\ell)^{t+1},\\
\end{equation}  
where $N_a(0)=1+\ell$ is chosen. 
The definition of this model is illustrated schematically in Fig.1.
 
\subsubsection{The multiplication rule}
In the case of the multiplication rule, the number of offsprings 
generated from each old vertex is not the same, but it depends on 
the degree of each vertex. 
Let $k_{i,m}(t)$ be the degree of vertex $i$ at time $t$, where 
the subscript $m$ means the multiplication rule. Then the number 
of offsprings generated at time $t+1$ from the vertex $i$ is proportional 
to its degree at the previous time, i.e., $(m-1)k_{i,m}(t)$, where 
$m-1$ is a proportionality constant. Thus the degree of vertex $i$ 
increases by a factor $m$ for each time step, that is, 
\begin{equation}
k_{i,m}(t)=m k_{i,m}(t-1). 
\end{equation}
Thus, the degree of vertex $i$ at 
time $t$ is  
\begin{equation}
k_{i,m}=m^{t-t_i}, 
\label{discrete}
\end{equation} 
for $t \ge t_i$. 
The total number of vertices newly born at time $t$, ${\cal {L}}_m(t)$ 
can be obtained to be 
\begin{eqnarray}
{\cal {L}}_m(t)&=&2(m-1)m\sum_{p=0}^{t-1}{t-1\choose p}
m^p (m-1)^{t-1-p},\nonumber \\
&=& 2m(m-1)(2m-1)^{t-1},  
\label{lmt}
\end{eqnarray}
for $t \ge 1$. The total number of vertices $N_m(t)$ at 
time $t$ is given by
\begin{equation}
N_m(t)=\sum_{j=0}^{t}{\cal {L}}_m(t)=1+m(2m-1)^t.\\ 
\end{equation} 
The definition of this model is illustrated schematically in Fig.2.

One may write the rate equation for the degree in this multiplicative 
process with continuous time as 
\begin{equation}
\frac{\partial k_{i,m}}{\partial t}=(m-1)k_{i,m}.  
\label{m_rate}
\end{equation} 
It would be interesting to compare this rate equation with the one 
for the preferential attachment (PA), in which the degree $k_i$ of vertex 
$i$ evolves as  
\begin{equation}
\frac{\partial k_i}{\partial t}={\cal L}_m\frac{k_i}{\sum_j k_j},
\label{rate_degree}
\end{equation}
where ${\cal L}_m(t)$ means the total number of edges newly 
introduced at time $t$. Since the total number of the degree at time 
$t$ is given by 
\begin{equation}
{\sum_{j} k_j} = 2m(2m-1)^{t-1}, 
\label{total_edge}
\end{equation}
and ${\cal {L}}_m$ is given by Eq.(\ref{lmt}), 
Eq.(\ref{rate_degree}) is reduced to Eq.(\ref{m_rate}), indicating that 
the rate equation for the degree in the multiplicative process is 
equivalent to the one for the preferential attachment.

\subsection{Loop structure}

The loop structure can be formed in networks, when a newly born vertex is 
connected to more than one existing vertices. For the loop structure, 
each already existing vertex generates the same number of offsprings 
as those for the tree structure. However, a newly born vertex is 
connected to two distinct old vertices: one is its parent, and the 
other is its grandparent. 
When the parent is one of vertices on $m$ branches (the centered 
one) born at $t=0$, the centered one (one of vertices on $m$ 
branches in a symmetrical way) is regarded as a grandparent. 
This rule is valid for both cases of the addition rule and the 
multiplication rule. The details on the connection rule is 
illustrated in Fig.3.   

\section{Analytic solution} 
\label{sec4}
\subsection{The degree distribution for the tree structure}
Since the degree of a vertex has been obtained explicitly as in 
Eqs.(\ref{kia}) and (\ref{discrete}) and they are ordered 
with time, we can obtain the degree distribution using the relation, 
\begin{equation}
P(k) =\frac{ \partial [1-P(k_{i}(t) > k)]}{\partial k},
\label{rel}
\end{equation}
which is valid for both cases of the addition and the multiplication rules. 
The detail of analytic treatments for the degree distributions for both 
cases are given as follows.  

\subsubsection{The addition rule}
Using the fact, $P_{a,t}(k_{i,a} > k)=P_{a,t}(t_i < t-(k-1)/\ell)$, 
where the subscript $t$ means the tree structure, we obtain that 
\begin{eqnarray}
P_{a,t}(k_{i,a}(t)> k)&=&\frac{\ell}{(1+\ell)^{t+1}}
\sum_{t_i=0}^{t-(k-1)/\ell}(1+\ell)^{t_i} \nonumber\\
&=& (1+\ell)^{-(k-1)/\ell}-(1+\ell)^{-(t+1)}.
\label{accm_a}
\end{eqnarray}
Applying Eq.(\ref{rel}) to Eq.(\ref{accm_a}), 
we obtain the degree distribution to be  
\begin{equation}
P_{a,t}(k)\propto (1+\ell)^{-(k/\ell)}.
\label{deg_a}
\end{equation}
So, the degree distribution $P_{a,t}(k)$ in the addition rule 
decays exponentially with $k$. 

\subsubsection{The multiplication rule}
Since the degree $k_i$ has been obtained explicitly as a function 
of time in Eq.(\ref{discrete}), $P_{m,t}(k_i > k)$ is written as 
$P_{m,t}(k_i > k)=P_{m,t}(t_i<\tau )$, where $\tau=t-\ln k/\ln m$. 
Since $P_{m,t}(t_i < \tau)$ means the density of the vertices born 
earlier than $\tau$, 
\begin{eqnarray}
P_{m,t}(k_i> k) &=& \sum_{t_i=0}^{\tau} \frac{{\cal {L}}_m(t_i)}{N_m(t)} 
\nonumber \\ &=& \sum_{t_i=0}^{\tau} {{2(m-1)(2m-1)^{t_i-1}}\over 
{1+m(2m-1)^t}}\nonumber \\
&\propto & k^{-\ln(2m-1)/\ln m}. 
\label{tree_pk}
\end{eqnarray}
Thus the degree distribution is obtained to be 
\begin{eqnarray}
P_{m,t}(k) &=& \frac{ \partial [1-P_{m,t}(k_{i}(t) > k)]}
{\partial k}\nonumber \\
       &\propto& k^{-\gamma(m)},
\end{eqnarray}
where 
\begin{equation}
\gamma(m)=1+\ln(2m-1)/\ln m.
\label{gamma}
\end{equation}
In the limit of $m \rightarrow 1$, we get $\gamma(1) =3$, while 
as $m$ goes to infinity, we get $\gamma (\infty)=2$. Thus by tuning 
the parameter $m$, we can get a variety of SF networks with 
different exponents in the range, $2 < \gamma < 3$.\\  

\subsection{The degree distribution for the loop structure}

\subsubsection{The addition rule}

Let $n_{i,a}(t)$ be the degree of vertex $i$ at time $t$ 
for the loop structure, where $a$ means the addition rule. 
Each old vertex receives edges from its $\ell$ children and 
$\ell^2$ grandchildren as they are born. So, Eq.(\ref{a_rate}) 
is modified as 
\begin{equation}
n_{i,a}(t+1)=n_{i,a}(t)+(\ell+\ell^2).
\end{equation}
Thus, the degree distribution shows an exponential-decay behavior,  
\begin{equation}
P_{a,l}(n)\propto (1+\ell+\ell^2)^{-n/(\ell+\ell^2)}, 
\end{equation}
where the subscript $l$ means the loop structure. 

\subsubsection{The multiplication rule}
Let $n_{i,m}(t)$ be the degree of vertex $i$ at time $t$ in the 
multiplication rule for the loop structure. The degree of vertex 
$i$ can be obtained, 
\begin{eqnarray}
n_{i,m}(t) &=&n_{i,m}(t-1)+(m-1)k_{i,m}(t-1)\cr 
&+&(m-1)^2 k_{i,m}(t-2), 
\end{eqnarray}
where the second term on the right hand side of the above equation 
results from the children of the vertex $i$, and the third term 
from the grandchildren of the vertex $i$. 
Thus, the degree at the vertex $i$ becomes 
\begin{equation}
n_{i,m}(t)=2m^{t-t_i}-m^{t-t_i-1}-m \approx \big(\frac{2m-1}{m}\big) m^{t-t_i}.
\end{equation}
Since the degree $n_{i,m}(t)$ depends on time $t$ similarly to 
Eq.(\ref{discrete}), we can apply Eq.(\ref{tree_pk}) even to the loop 
case, except that $\tau$ is replaced by 
$\tau=t+\ln(2m-1)/\ln m -1-\ln n /\ln m$. 
This replacement, however, does not affect the degree exponent at all. Thus, 
even for the loop structure, the degree exponent is reduced to the same 
value, $\gamma=1+\ln(2m-1)/\ln m$ in Eq.(\ref{gamma}).  

\subsection{The diameter for the tree structure}
The diameter $d(t)$ is defined as a geodesic distance between two distinct 
vertices along the shortest path averaged over all pairs of vertices 
at time $t$, that is, 
\begin{equation}
d(t) =\frac{1}{N(t)(N(t)-1)} \sum_{i \neq j} d_{i,j}(t), \label{eq:64}
\label{dia}
\end{equation}
where $d_{i,j}(t)$ is the shortest-path distance between vertex 
$i$ to $j$. For simplicity, let ${\cal {D}}(t)$ denote the sum 
of the shortest-path distances between two vertices over all pairs, 
that is,  
\begin{equation}
{\cal D}(t)=\sum_{i \neq j}d_{i,j}(t). 
\end{equation}
It is not easy to obtain a closed formula for ${\cal D}(t)$ 
for both the tree and the loop structure, however, 
we list ${\cal D}(t)$ for the tree structure in a few early times 
in Appendix. We trace the formula for the tree structure 
in two limiting cases, $m\rightarrow 0$ and $m\rightarrow \infty$, 
as follows.\\
 
Let us first consider the case of $m \rightarrow 1$. 
For this case, we denote $m=1+\epsilon $ and $\epsilon \ll 1$. 
The total number of nodes $N(t)$ at time $t$ is given by
\begin{eqnarray}
N(t)&=& 1+(1+2\epsilon)^t (1+\epsilon) \nonumber \\
&\approx& 2+(2t+1)\epsilon + {\cal{O}}(\epsilon^2). 
\end{eqnarray}
Moreover, the sum of all shortest-path distances ${\cal D}(t)$ becomes 
\begin{equation}
{\cal {D}}(t)\approx 2+4(2t+1)\epsilon+{\cal O}(\epsilon^2).
\end{equation}
Using the relation Eqs.(\ref{dia}), we can obtain the average 
distance to be 
\begin{eqnarray}
d &=& \frac{2+4(2t+1)\epsilon}{2+3(2t+1)\epsilon} + 
{\cal{O}}(\epsilon^2), \nonumber \\
&\cong & \frac{-8}{7+6\log (N-1)} +\frac{4}{3}. \label{eq:73}
\end{eqnarray}
Therefore, the diameter converges to $4/3$ in the limit 
of $N \rightarrow \infty$.\\

Next, we consider the case of $m\rightarrow \infty$. In this case, 
the term in the highest order of $m$ could be dominant, so that 
we trace the coefficient of the term in the highest order of $m$ as a 
function of time.  
\begin{eqnarray}
{\cald}(0)&=& 2m^2+\mathrm{lower~order~terms} \nonumber \\
{\cald}(1)&=& [(2+3)+(3+4)]m^4 + \mathrm{lower~order~terms} \nonumber \\
{\cald}(2)&=& [(2+2\cdot 3+4)+2\cdot (3+2\cdot 4+5) +(4+2\cdot 5+6)]m^6
              \nonumber \\
           && +\mathrm{lower~order~terms} \nonumber \\
{\cald}(3)&=& [(2+3\cdot 3+3\cdot 4+5)+3(3+3\cdot 4+3\cdot 5+6) \nonumber \\
&& 3(4+3\cdot 5+3\cdot 6+7)+(5+3\cdot 6+3\cdot 7+8)]m^8 \nonumber \\
&& +\mathrm{lower~order~terms} \nonumber \\
     && \vdots \nonumber \\
{\cald}(t) &=& \sum_{p=0}^{t}{t\choose p}\sum_{k=0}^{t}{t\choose k}
              (k+p+2)m^{2(t+1)} \nonumber \\
           && +\mathrm{lower~order~terms}. 
\end{eqnarray}
Therefore the term in the highest order of $m$ for ${\cald}(t)$ is 
obtained explicitly to be 
\begin{equation}
{\cal {D}}(t)\approx(t+2)2^{2t}m^{2(t+1)}, \label{eq:78} 
\end{equation}
where the coefficient $(t+2)2^{2t}$ means the number of pathways 
having the distance $2(t+1)$, which is the farthest one at time $t$ 
in the system. On the other hand, 
\begin{eqnarray}
N(t)(N(t)-1) &\approx& 2^{2t} m^{2(t+1)}.  
\end{eqnarray}
Therefore, the diameter $d(t)$ at time $t$ becomes simply 
\begin{eqnarray}
d(t) &\approx & t+2 \nonumber \\
   &\approx &\frac{\log(N-1)}{\log (2m-1)}+2.
\label{eq:79}
\end{eqnarray}
Thus, for large $N$, the above equation is rewritten simply as 
\begin{equation}
d(N)\sim \ln N/\ln {\bar k}
\end{equation} 
with the mean degree ${\bar k}\approx 2m$, which confirms the small-world 
behavior.   

\section{Conclusions and Discussions}
We have introduced a deterministic model for the scale-free network, 
which is constructed in a hierarchical way. At each time step, 
each already existing vertex produces its offsprings, whose number 
is proportional to the degree of the vertex. Depending on whether each 
offspring is connected to only one or more than one old vertices, 
the network forms either a tree structure or a loop structure, 
respectively. We have obtained the analytic solution for the degree 
distribution and the diameter explicitly for the deterministic 
model. By tuning a control parameter in the model, we can adjust 
the degree exponent in the range, $2 < \gamma < 3$. Thus this model 
can represent a variety of SF networks in real world. 
Moreover, we obtained the diameter of the deterministic model 
analytically to be $d \sim \ln N/\ln {\bar k}$, where $N$ is the system 
size and $\bar k$ is the mean degree. Since the network is generated 
in a hierarchical way, it is expected that a variety of physical 
problems can be solved through this deterministic model by 
constructing recursive relations derived from two structures in 
successive generations. On the other hand, the deterministic model 
has a shortcoming that it does not include any long-ranged edge, 
connecting two vertices belonging to different branches separated 
at $t=0$. Thus, while this model can be easily generalized by 
controlling the number of branches $m$, it is extremely vulnerable, 
and can be broken into pieces by a simple deletion of the centered vertex.   
Despite this shortcoming, we think that our deterministic model could 
offer a guide toward generating more realistic deterministic model 
for SF networks.  
 
\section{Acknowledgment} 
We acknowledge the supports by the Ministry of Science \& Technology 
through the National Reseach Lab. program, and by grants 
No. R01-2000-00023 from the BRP program of the KOSEF.

\section{Appendix}
The closed formula for the sum of the shortest-path distance between 
two vertices, ${\cald}(t)$ are shown for $0 \le t \le 2$.  
\begin{eqnarray}
{\cald}(t=0) &=& {\cal N}_0 m+{\cal N}_{0,0}[1+2(m-1)], \nonumber \\
{\cald}(t=1) &=& {\cal N}_0 [m^2+2m(m-1)]+{\cal N}_{0,0} 
[m+2m(m-1)+2(m-1)+3(m-1)(m-1)]\nonumber \\
        &+&{\cal N}_{1,0}[1+2m^2-2(m+1)+2m+3m(m-1)] \nonumber \\
        &+&{\cal N}_{1,0,0}[1+2(m-1)+3(m^2-1)+4(m-1)^2],\nonumber \\
{\cald}(t=2) &=& {\cal N}_0[m^3+2(m^2-m)(m-1)+2m(m-1)+2m(m^2-m)
        +3m(m-1)(m-1)]\nonumber \\
        &+& {\cal N}_{0,0}[1m^2+2m^3-2+2(m-1)(m-1)+3\{(m-1)^2+(m^2-m)
        (m-1)\nonumber \\
        &+&(m-1)(m^2-m)\}+4(m-1)^3] \nonumber \\
        &+& {\cal N}_{1,0}[m+2(m^3-1)+3\{(m(m-1)-1)(m-1)+m(m^2-m)+m(m-1)\}
\nonumber \\
        &+&4m(m-1)^2 ] \nonumber \\
        &+& {\cal N}_{1,0,0}[m+2(m^2-1)+3\{(m^3-1)+(m-1)(m-2)\}
\nonumber \\
        &~~&+4\{(m^2-m)(m-1)+(m-1)^2+(m-1)(m^2-m)\}+5(m-1)^3] \nonumber \\
        &+&{\cal N}_{2,0}[1+2(m^3-1)+3\{m^2(m-1)+m(m^2-m)\}+4m(m-1)^2] \nonumber \\
        &+&{\cal N}_{2,1,0}[1+2(m-1)+3(m^3-1)+4\{(m(m-1)-1)(m-1)+
m(m^2-1)\}\nonumber \\ 
&+&5m(m-1)^2] \nonumber \\
        &+&{\cal N}_{2,0,0}[1+2(m^2-1)+3\{(m^3-1)+(m-1)(m-1)\} 
\nonumber \\
        &~~& +4\{(m^2-1)(m-1)+(m-1)(m^2-m)\}+5(m-1)^3] \nonumber \\
        &+&{\cal N}_{2,1,0,0}[1+2(m-1)+3(m^2-1)+
4\{(m^3-1)+(m-1)(m-1-1)\} \nonumber \\
        &~~&+5\{(m^2-1)(m-1)+(m-1)(m^2-1)\}+6(m-1)^3], \nonumber \\
        &  & \vdots  \nonumber 
\end{eqnarray}
where 
\begin{eqnarray}
  &&{\cal N}_{0}= 1, \cr 
  &&{\cal N}_{0,0} = m, \cr 
  &&{\cal N}_{1,0} = m^2-m, \cr 
  &&{\cal N}_{1,0,0} = m(m-1), \cr 
  &&{\cal N}_{2,0} = m^3-m^2, \cr 
  &&{\cal N}_{2,1,0}= (m^2-m)(m-1), \cr 
  &&{\cal N}_{2,0,0} = m(m^2-m), \cr 
\noalign{\hbox{and}}
  &&{\cal N}_{2,1,0,0} = m(m-1)(m-1).\nonumber  
\end{eqnarray}
${\cal N}_{i,j}$ means the number of the vertices denoted by $A_{i,j}$ 
in Fig.2, where the first index $i$ stands for its birth time and 
the rest indices $j$ do its parent vertex.

\begin{figure}
\centerline{\epsfxsize=8cm \epsfbox{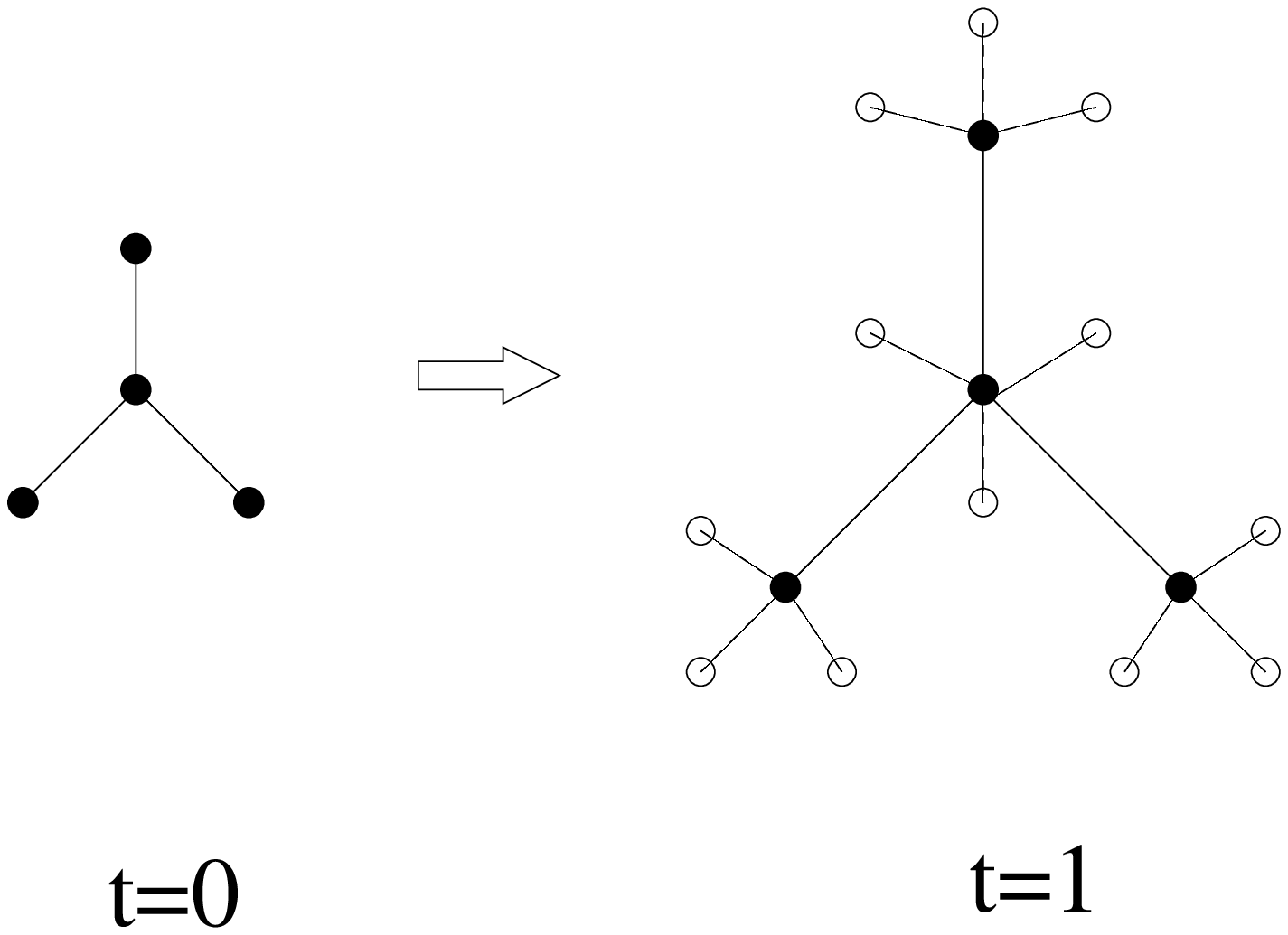}}
\caption{Tree structures in the addition rule with $\ell=3$ 
at $t=0$ and $t=1$.} 
\label{fig:add}
\end{figure}

\begin{figure}
\centerline{\epsfxsize=8cm \epsfbox{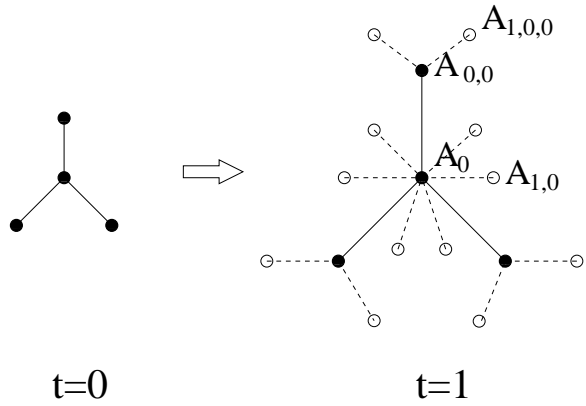}}
\caption{Tree structures in the multiplication rule with $m=3$ 
at $t=0$ and $t=1$. $\rm A_0$ stands for the vertex at center, 
$\rm A_{0,0}$, a neighbor of $\rm A_0$ born at $t=0$, and 
$\rm A_{1,0}$ ($\rm A_{1,0,0}$), an offspring of $\rm A_0$ 
($\rm A_{0,0}$) born at $t=1$.}
\label{fig:mul}
\end{figure}

\begin{figure}
\centerline{\epsfxsize=8cm \epsfbox{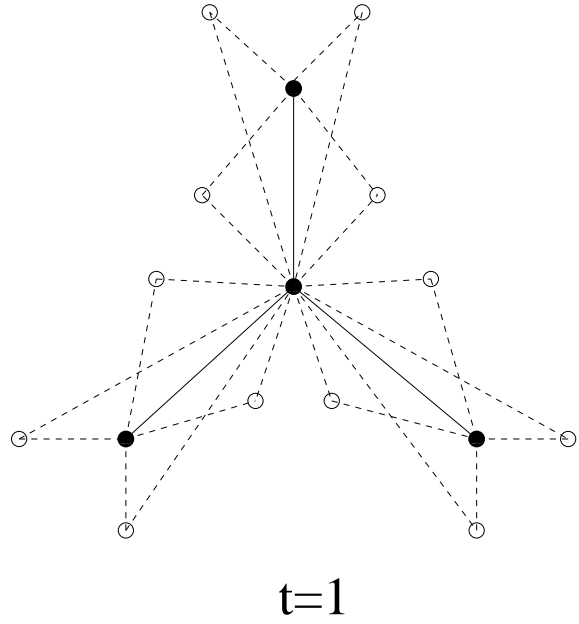}}
\caption{Loop structure in the multiplication rule with $m=3$ 
at $t=1$.}
\label{fig:mul}
\end{figure}

\end{document}